\documentclass[12pt,preprint,a4paper]{aastex}
\usepackage{graphics,epsf}
\usepackage{amsmath}                
\usepackage{amsfonts}               
\usepackage{amssymb}                
\usepackage{epsfig}                 
\usepackage{subfigure}
\usepackage{float}

\def \cm{~\rm{cm}}
\def \s{~\rm{s}}
\def \km{~\rm{km}}

\def \K{~\rm{K}}
\def \g{~\rm{g}}

\def \erg{~\rm{erg}}

\def \yr{~\rm{yr}}

\begin{document}

\title{The response of a helium white dwarf to an exploding type Ia supernova}

\author{Oded Papish\altaffilmark{1}, Noam Soker\altaffilmark{1}, Enrique Garc\'\i a--Berro\altaffilmark{2,3}, and Gabriela Aznar--Sigu\'an\altaffilmark{2,3}}

\altaffiltext{1}{Department of Physics, Technion -- Israel
Institute of Technology, Haifa 32000, Israel;
papish@tx.technion.ac.il; soker@physics.technion.ac.il.}
\altaffiltext{2}{Departament de F\'\i sica Aplicada, Universitat
Polit\`ecnica de Catalunya, c/Esteve Terrades 5, 08860
Castelldefels, Spain} \altaffiltext{3}{Institute for Space
Studies of Catalonia,
                c/Gran Capit\`a 2--4, Edif. Nexus 201,
                08034  Barcelona, Spain}

\begin{abstract}
We conduct numerical simulations of the interacting ejecta from
an exploding CO white dwarf (WD) with a He~WD donor in the
double-detonation scenario for Type Ia supernovae (SNe Ia), and study the possibility of exploding the companion WD. We also study the long time imprint of the collision on the supernova remnant. When the donor He~WD has a low mass, $M_{\rm WD} =0.2 M_\odot$, it is at a distance of $\sim 0.08 R_\odot$ from the explosion, and  helium is not ignited. The low mass He~WD casts an `ejecta shadow' behind it. By evolving the ejecta for longer times, we find that the outer
parts of the shadowed side are fainter and its boundary with the
ambient gas is somewhat flat. More massive He~WD donors, $M_{\rm WD} \simeq 0.4M_\odot$, must be closer to the CO~WD to transfer mass. At a
distance  of $a \la 0.045 R_\odot$  helium is detonated and the
He~WD explodes, leading to a triple detonation scenario. In the explosion of the donor WD approximately $0.15M_\odot$ of unburned helium is ejected. This {{{{ might be observed}}}} as a peculiar type Ib supernova.
\end{abstract}
\keywords{ISM: supernova remnants --- supernovae: stars: binary
 --- binaries: close  --- hydrodynamics ---
supernovae: general}

\section{INTRODUCTION}
\label{sec:intro}

Type Ia Supernovae (SNe Ia) are one of the most energetic events
in the universe, now known to be originated by thermonuclear
detonations of carbon-oxygen (CO) white dwarfs
(\citealt{Hoyle1960}). Several possible scenarios leading to a SN
Ia outburst are currently envisaged, although there might be some
overlap between them.  All scenarios have advantages and drawbacks
(e.g., \citealt{TsebrenkoSoker2015b}), and there is not yet a
general consensus on the leading scenario for SN Ia. In fact, it
is well possible that all of them contribute to the total SN Ia
rate in some unknown fraction.

 These scenarios can be listed as follows, according to alphabetical order.
(a){\it The core-degenerate (CD) scenario} (e.g.,
\citealt{Livio2003, Kashi2011, Soker2011, Ilkov2012, Ilkov2013,
Soker2013, TsebrenkoSoker2015a}). Within this scenario the WD
merges with the hot core of a massive asymptotic giant branch
(AGB) star. In this case the explosion might occur shortly or a
long time after the merger. In a recent paper, \cite{TsebrenkoSoker2015b} argue
that at least $20 \%$, and likely many more, of all
SNe Ia come from the CD scenario. (b){\it The double degenerate
(DD) scenario} (e.g., \citealt{Webbink1984, Iben1984}). This
scenario is based on the merger of two WDs. However, this scenario
does not specify the subsequent evolution of the merger product, namely,
how long after
the merger the explosion of the remnant takes place (e.g.,
\citealt{vanKerkwijk2010}). Recent papers, for example, discuss
violent mergers (e.g., \citealt{Loren2010, Pakmoretal2013})
as possible channels of the DD scenario, while others
consider very long delays from merger to explosion, e.g., because
rapid rotation keeps the structure overstable
\citep{TornambPiersanti2013}. \cite{Levanonetal2015} argue that the
delay between merger and explosion in the DD scenario should be
$\gg 10 \yr$. (c){\it The single degenerate (SD) scenario} (e.g.,
\citealt{Whelan1973, Nomoto1982, Han2004}). In this scenario a
white dwarf (WD) accretes mass from a non-degenerate stellar
companion and explodes when its mass reaches the Chandrasekhar
mass limit. (d){\it The double-detonation mechanism} (e.g.,
\citealt{Woosley1994, Livne1995}). Here a sub-Chandrasekhar mass
WD accumulates a layer of helium-rich material coming from a helium donor
on its surface. The helium layer is compressed as more material is
accreted and detonates, leading to a second detonation near the
center of the CO WD (see, for instance, \citealt{Shenetal2013} and
references therein, for a recent paper). (e) \emph{The WD-WD
collision scenario} (e.g., \citealt{Thompson2011, KatzDong2012,
Kushniretal2013, Aznar2014}). In this scenario either a tertiary star brings two WDs
to collide, or the dynamical interaction occurs in a dense stellar system, where such interactions are likely. In some cases, the collision results in an immediate explosion. Despite
some attractive features of this scenario, it can account for at
most few per cent of all SNe Ia \citep{Hamersetal2013,
Prodanetal2013, Sokeretal2014}.

Finally, it should be mentioned that very recently it has been
suggested that pycnonuclear reactions could be able to drive
powerful detonations in single CO white dwarfs
\citep{Chiosietal2014}. This scenario -- the so-called {\it single
WD scenario} -- has, however, two important shortcomings. The
first one is that the typical H mass fraction found in detailed
evolutionary calculations of CO WD progenitors is much smaller
than that needed to ignite the core of the WD. The second drawback
of this recently suggested scenario is that most SN Ia come from
WDs with masses near the Chandrasekhar limit  (e.g.,
\citealt{Seitenzahletal2013, Scalzoetal2014}), while the mass at
which ignition may possibly occur in the single WD scenario is
$\sim 1.2 M_\odot$.  Hence, this scenario  would also only account
for a small percentage of all SN Ia.

As mentioned earlier, there is some overlap between these
scenarios. For example, in the violent merger model
\citep{Loren2009, Pakmor2012} it is possible that during the first
stages of the merger of the two CO WDs the small helium buffer
($\simeq 10^{-2}\, M_{\sun}$) of the original CO WDs is ignited.
In this case  both the DD scenario and the double detonation
mechanism operate simultaneously. Also, the double detonation
mechanism might operate in the CD scenario.

In this  paper we study the response of a donor star that is a He
WD to an exploding CO WD with mass below the Chandrasekhar limit,
$M_{\rm WD} \simeq 1.0-1.1 M_\odot$. These parameters fit the
double detonation scenario where a very low mass helium shell
triggers the SN Ia explosion of a CO WD \citep{Bildstenetal2007,
ShenBildsten2009, ShenBildsten2014}.  We will answer five
questions. (1) Does the shock wave induced by the ejecta ignite
helium in the WD companion by adiabatic compression or by shock
heating? (2) Is carbon in the ejecta ignited as it is shocked in
the outer layers of the He WD?  (3) Can mixing of helium from the
donor and carbon from the ejecta lead to vigorous nuclear burning?
(4) How much helium is entrained by the ejecta? (5) What is the
morphology of the SNR long time after the explosion as the SN
ejecta sweep some ambient medium gas? To do so we will adopt two
masses for the He WD companion. First we study analytically and
then numerically the impact of the  SN Ia ejecta of a WD of mass
$0.43 M_\odot$ residing at $\sim 0.02-0.03 R_\odot$ from the
exploding CO WD. This setting is based on the numerical
simulations of \cite{Guillochonetal2010}, \cite{Raskin et al2012},
and \cite{Pakmoretal2013}, for similar (but not identical)
progenitors that might lead to SN Ia. In a second step, and
following \cite{Bildstenetal2007} and \cite{ShenBildsten2009} we
also consider a He WD of $0.2M_\odot$ at an orbital separation of
$0.08 R_\odot$.

 { {{{ There are a number of simulations
studying similar processes to those studied by us, but in the SD
scenario.  \cite{Mariettaetal2000} conducted 2D simulations to
study the impact of a SN Ia on a hydrogen-rich non-degenerate
companion. They found that several tenths of a solar mass of
hydrogen are striped from the companion into a cone with a solid
angle of $65-115 ^\circ$ behind the companion, depending on the
type of companion. \cite{Kasen2010} was interested in the effect of the
companion on the light curve shortly, up to several days, after
the explosion.  \cite{Pakmoretal2008} found the striped
hydrogen mass to be much lower, and compatible with limits from
observations. \cite{Panetal2010} took the companion in their 2D
simulations to be a non-degenerate helium star.
\cite{Panetal2012a} extended the study to 3D simulations and to
hydrogen-rich companions. \cite{Panetal2012b} were interested in
the evolution of a main sequence companion after the passage of
the SN shock. We reproduce the dense conical surface found to be
formed behind the companion by \cite{Panetal2010} and
\cite{Panetal2012a}, but we continue to follow the interaction of
this cone with the ISM. We note that none of the papers listed
above continued their simulations to the stage of interaction with
the ISM, as we do in the present study. Neither they included
nuclear reactions in the companion as a result of the shock. Here
we study the interaction of a type Ia supernova with a He~WD to
examine He ignition and the SNR morphology.}}}}

Our paper is organized as follows. In section \ref{sec:ejecta} we
discuss and quantify the properties of the material ejected from
the disrupted CO WD, while in section \ref{sec:ignition} we assess
analytically  the possibility of an explosive ignition. In section
\ref{sec:numeric} we conduct 2D axisymmetrical numerical
simulations of the interaction of the ejecta with the He WD, and
we examine nuclear reactions and helium entrainment. Finally, in section
\ref{sec:conclusions} we summarize our results and their
implications to the double detonation scenario.

\section{EJECTA PROPERTIES}
\label{sec:ejecta}

To facilitate an analytical estimate we assume that the SN Ia
ejecta is already in homologous expansion, and we take the profile
of \cite{Dwarkadas1998}
\begin{equation}
\rho_{\rm{SN}} = A \,{\rm{exp}}{} (-v/v_{\rm{ejecta}})t^{-3},
\label{eq:rhosn}
\end{equation}
where $v_{\rm{ejecta}}$ is a constant which depends on the mass and kinetic energy of the ejecta,
\begin{equation}
v_{\rm{ejecta}} = 2.9 \times 10^8 E_{51}^{1/2}
\left(\frac{M_{\rm{SN}}}{1 M_\odot}\right)^{-1/2} \cm \s^{-1},
\label{eq:vejacta}
\end{equation}
$E_{51}$ is the explosion energy in units of $10^{51} \erg$, and
$A$ is a parameter given by
\begin{equation}
A = 3.3 \times 10^6 \left(\frac{M_{\rm{SN}}}{1
M_\odot}\right)^{5/2} E_{51}^{-3/2}  \g \s^{3} \cm^{-3} .
\label{eq:aconst}
\end{equation}
The maximum velocity of the SN Ia ejecta is $v_{\rm SNm} \simeq
20,000 \km \s^{-1}$. { {{{ We compared this analytical
profile with $M_{\rm{SN}}=1 M_\odot$ and
 $E_{51}=1$ with models 7D and 9C from \cite{WoosleyKasen2011}, who calculated the explosion of WD models.
The maximum velocity in the analytical profile used here is
$20,000 \km \s^{-1}$. We found our model to be practically identical to  their model 7D for the outer $0.2 M_\odot$ of the ejecta, and somewhat
slower than model 9D in that mass range. For inner mass
coordinates the analytical fit is slower than models 7D and 9C of
\cite{WoosleyKasen2011}. As the outer layers determine whether the
companion will be ignited, using models 7D or 9C of
\cite{WoosleyKasen2011} will result in an easier ignition of the
companion.
  For that, and to keep the profile simple and flexible to changes,
}}}}  we use the analytical profile as given above both in the
analytical and the numerical calculations.

For the analytical estimates derived in section \ref{sec:ignition}
we now estimate the maximum ram pressure of the ejecta on the
He~WD. A cold He WD of mass $0.43\, M_{\sun}$ has a radius of
$\sim 0.015R_\odot$. As it overflows its Roche lobe, with a CO~WD
companion of $1 M_\odot$, in a stable mass transfer the orbital
separation is $\sim 3.3$ times this distance, namely, $a \simeq
0.05 R_\odot$. However, detailed numerical calculations show that
for a powerful ignition to occur the mass transfer must be unstable
\citep{Guillochonetal2010}, and the surface of the He~WD that
fills the Roche lobe can be as close as $\sim 0.02 R_\odot$ to the
exploding CO~WD \citep{Raskin et al2012, Pakmoretal2013}.

The ram pressure of the ejecta at a distance $r_{\rm e}$ from the
explosion at time $t$ after explosion is given by
\begin{equation}
P_{\rm ram}=\rho(r_{\rm e}) v^2 =
A \,{\rm{exp}}{} (-r_{\rm e}/v_{\rm{ejecta}}t)t^{-5} r_{\rm e}^2,
\label{eq:pram1}
\end{equation}
where $v=r_{\rm e}/t$. The maximum ram pressure is achieved at
time $t_{\rm max}=r_{\rm e}/(5 v_{\rm{ejecta}})=1 (r_{\rm
e}/0.02\, R_{\sun}) \s$, and its value is
\begin{equation}
P_{\rm ram}^{\rm max}=5.2 \times 10^{22}
E_{51} \left(\frac{r_{\rm e}}{0.02\, R_{\sun}} \right)^{-3} \erg \cm^{-3}.
\label{eq:pramm}
\end{equation}
At $t=2t_{\rm max}$ and $t=3 t_{\rm max}$ the pressure drops to a
value of $0.38 P_{\rm ram}^{\rm max}$ and $0.12 P_{\rm ram}^{\rm
max}$, respectively. The first material hits the WD at time $\sim
0.02\, R_{\sun} / 20,000$~km ~s$^{-1} = 0.7 \s \simeq 0.7 t_{\rm
max}$, with a ram pressure of $0.7 P_{\rm ram}^{\rm max}$.
Overall, the phase in which the pressure is larger than  $\sim 0.3
P_{\rm ram}^{\rm max}$ lasts for about two seconds at $\sim 0.02
R_\odot$ from the explosion. The density of the ejecta at maximum
ram pressure is
\begin{equation}
\rho(t_{\rm max}) = 2.5 \times 10^4 \left(\frac{M_{\rm{SN}}}{1
M_\odot}\right) \left(\frac{r_{\rm e}}{0.02\, R_{\sun}}
\right)^{-3} \g \cm^{-3} .
   \label{eq:rhoej}
\end{equation}

\section{CONDITIONS FOR NUCLEAR IGNITION}
\label{sec:ignition}

Fig.~\ref{f:profilePrho} shows two of the physical quantities of a
$0.43\, M_{\sun}$ He WD which are relevant for our study, namely the
pressure and density as a function of the mass coordinate
$-\log(1-M_r/M_{\rm WD})$. This specific model corresponds to a WD
with central temperature $T\simeq 10^7$~K, which results in a
surface luminosity $\log(L/_{\sun})\sim -2.85$, an otherwise
typical luminosity of field white dwarfs, an effective temperature
$\log T_{\rm eff}\simeq 3.93$, and corresponds to a sequence which
was evolved performing full evolutionary calculations that
consider the main energy sources and processes of chemical
abundance changes during white dwarf evolution
\citep{Althaus2009}. There are three possible ways in which the He~WD or the CO ejecta
might be ignited:

\begin{figure}[t]
   \resizebox{\hsize}{!}
{\includegraphics[width=\columnwidth]{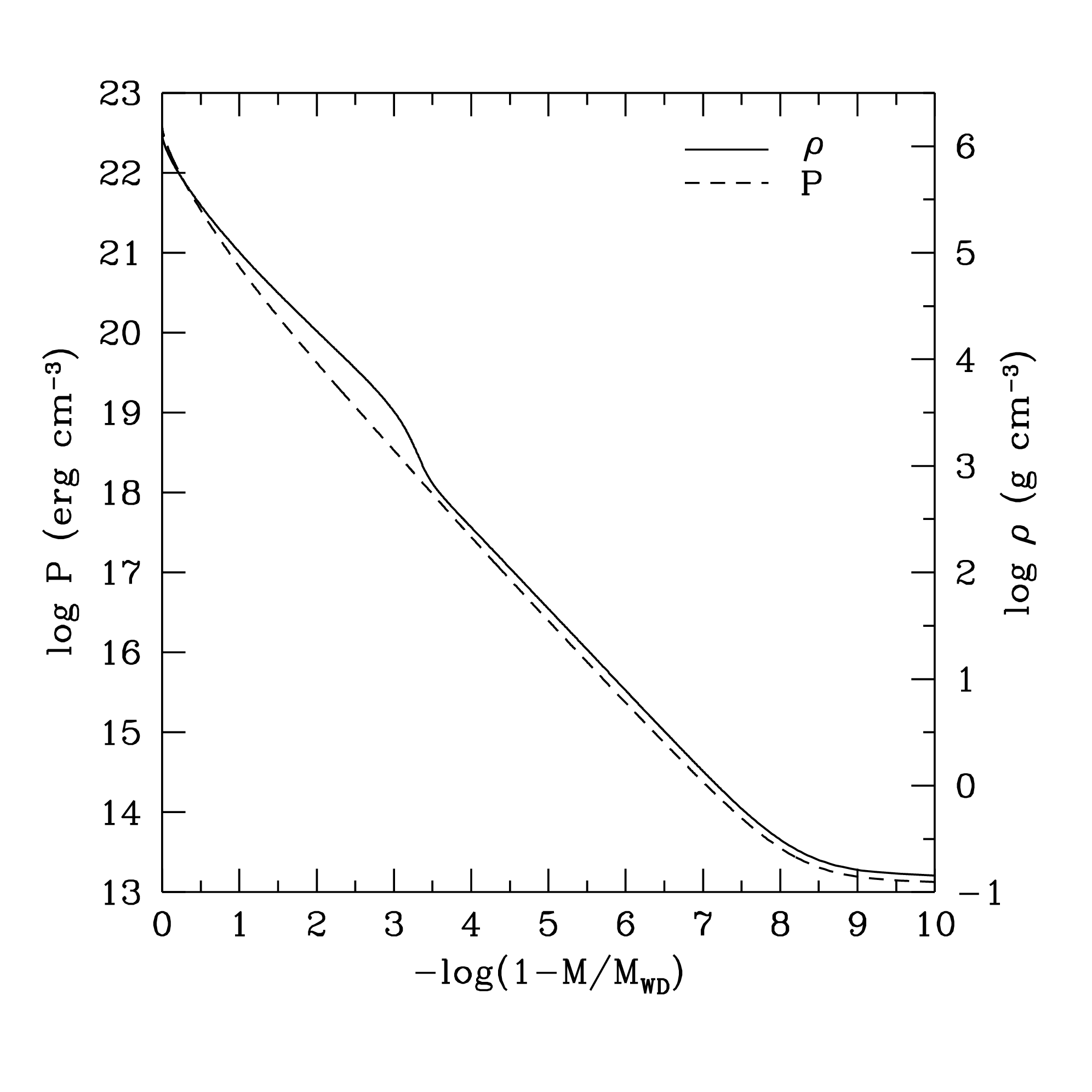}}
\caption{Pressure and density profiles of a $0.43\, M_{\sun}$ He
WD, as a function of the mass coordinate $\log(1-M_r/M_{\rm WD})$.
This coordinate allows to better resolve the very outer layers of
the star, where the effects of the shock are presumably more
important. The central temperature of the WD is $10^7 \K$.}
\label{f:profilePrho}
\end{figure}

{(1) \it Shock ignition of helium.}  It  turns out that, for the
model WD used here, He is shocked and ignited in a region where
 both thermal and radiation pressures play a role. In this region $\rho
\sim 10^5 \g \cm^{-3}$ and $T \simeq 1.2 \times 10^9 \K$. A good
estimate of the temperature in the shocked region of the He~WD can be
obtained by equating the radiation pressure to the ram pressure
given in equation (\ref{eq:pramm}):
\begin{equation}
T_{\rm He} \simeq 1.2 \times 10^9
    \left(\frac{r_{\rm e}}{0.04 R_{\sun}} \right)^{-3/4}
      \K.
\label{eq:TempHeWD}
\end{equation}
The burning time-scale of pure helium at these conditions is $\sim 10 \s$, just a little longer
than the timescale of the dynamical flow,  defined as the ejecta speed divided by
the He WD radius, $\sim 0.04 R_\odot /10,000 \km \s^{-1} \sim
3 \s$. For these parameters, ignition conditions are reached  for $r \la 0.04 R_\odot$.
This is compatible with the numerical
results to be described in section \ref{subsec:MassiveWD}, where
the exact radius is found.

{(2) \it Carbon burning in the shocked ejecta.} The second
possibility we explore is the ignition of carbon-rich material of
the ejecta as it is shocked upon hitting the He WD. The post-shock
pressure of the ejecta is dominated by radiation pressure. The
temperature at maximum ram pressure is given by equation
(\ref{eq:TempHeWD}). For a distance to the explosion $r_e = 0.02
R_\odot$ we find the temperature to be $T_{\rm CO} \simeq 2 \times
10^9 \K$. For this temperature we expect that carbon will be
burned. Nevertheless, we need to compare the burning time with the
dynamical timescale of the flow, $\tau_{\rm flow} \sim 1 \s$. For
the scaling and parameters used in Sect.~\ref{sec:ejecta} the
ejecta density at the time of maximum ram pressure and at a
distance of $0.02\, R_{\sun}$ from the center of explosion is $3.5
\times 10^4 \g \cm^{-3}$. If the carbon mass is half of the mass
of the ejecta and the compression factor is $\sim 4$, then the
post-shock density in the carbon-rich region is $\rho_{\rm C}
\simeq 7 \times 10^4 \g \cm^{-3} \sim 10^5 \g \cm^{-3}$. As in
this scenario the companion star is much closer to the center of
the explosion than the corresponding one of the single-degenerate
scenario, the density of the shocked ejecta will be much higher,
and the burning timescale much shorter. We find that the carbon
burning timescale for this density and a typical temperature $\sim
2 \times 10^9 \K$ to be about one second. These temperatures and
densities are achieved near the stagnation point in a small region
of  size $\sim 0.1 r_e$ -- see below. The outflow time from this region is
$\sim 0.002 R_\odot/1.5 \times 10^4 \km \s^{-1} = 0.1 \s$. Thus,
the outflow time is shorter than the burning time scale.  In the
numerical results to be described next we obtain no significant
carbon burning, showing that the outflow time of carbon from the
shocked region is indeed very short.  This  is unlike  the case in
which   helium   belonging to the  He~WD is shocked inside the
He~WD and cannot flow outward.

{\it (3) Igniting helium by mixing ejecta.} Even if carbon is not
ignited, mixing of the ejecta at $T \sim 10^9 \K$ with helium
might, in principle,  even if helium was not ignited by the shock,
power a thermonuclear runaway. {{{{  In our numerical
simulations mixing is not sufficiently deep to cause ignition
by this process (see section \ref{sec:numeric}). }}}}

For the case of a low-mass He~WD we repeated all these calculations
and we found that none of the previously described processes drive
a powerful nuclear outburst, and thus the evolution in this case
should mostly consist of a purely hydrodynamical flow. As it will be
explained in detail in the next section, full hydrodynamical numerical
simulations confirm this.

\section{NUMERICAL SIMULATIONS}
\label{sec:numeric}
\subsection{Numerical setup}
\label{subsec:setup}
We use version 4.2.2 of the FLASH gas-dynamical numerical code
\citep{Fryxell2000}. {{{{ The FLASH code has been used
before for a similar study in the SD scenario, in 2D
\citep{Kasen2010, Panetal2010} and 3D \citep{Panetal2012a,
Panetal2012b}.
 }}}}
 The widely used \texttt{FLASH} code is a publicly
available code for supersonic flow suitable for astrophysical
applications. The simulations are done using the unsplit PPM
solver of \texttt{FLASH}. We use 2D axisymmetric cylindrical coordinates
with an adaptive mesh refinement (AMR) grid. The origin of the
grid, $(0,0)$, is taken at the center of the explosion. In all the
figures shown below the symmetry axis of the grid is the vertical
axis. The axisymmetric grid forces us to neglect the orbital
relative velocity of the He~WD and the exploding CO~WD. In any
case, the orbital velocity is much smaller than the ejecta
velocity, and will have virtually no effect on our conclusions.
{ {{{ For the equation of state we use the Helmholtz EOS
\citep{TimmesSwesty2000}. This EOS includes
contributions from partial degenerate electrons and positrons, radiation, and non-degenerate ions. We use
the Aprox19 nuclear network of 19 isotopes \citep{Timmes1999} in \texttt{FLASH}.  The hydrodynamic module is coupled to the nuclear network by setting the parameter \texttt{enucDtFactor=0.1} in \texttt{FLASH} \citep{Hawleyetal2012}, and shock
burning is disabled by default. Self gravity is included using the new multipole solver in \texttt{FLASH} with order $l=10$.  }}} }

{{{ {We run our collision simulations with two different resolutions as a test for convergence and found no appreciable difference. In addition, we run a low resolution simulation on a much larger grid to follow the long time evolution of the ejecta. For the high-resolution simulations the minimum
cell size was $\sim12\times12 \km$ with a total of 10 levels of AMR
refinement. For the low-resolution simulations the minimum cell
size was $\sim48\times48 \km$. In addition we lowered the resolution in the large grid simulation
to  $\sim 92\times 92 \km$ from initially $\sim46\times46 \km$ after $t=16 \s$ from the explosion to reduce the computational time.} }}}

The initial He~WD mass, radius, and distance from the center of
the explosion in the two simulated cases to be presented below
are $(M_{\rm WD}, R_{\rm WD}, a_0)=(0.2 M_\odot, 0.02 R_\odot, 0.082 R_\odot)$ and
 $(M_{\rm WD}, R_{\rm WD}, a_0)=(0.43 M_\odot, 0.015 R_\odot,
0.029-0.043 R_\odot)$
 for the low- and high-mass He~WDs, respectively.
 The WDs are cold, and the radius of the $0.43 M_\odot$ WD is
 somewhat smaller than the hotter WD presented in
 Fig.~\ref{f:profilePrho}.
These models were built with version 6022 of the Modules for
Experiments in Stellar Astrophysics (MESA; \citealt{Paxton2011}).

Initially, the ejecta in our simulations is homologous expanding
according to equations (\ref{eq:rhosn})-(\ref{eq:aconst}), with
$E_{51}=1$ and $M_{\rm SN} = 1 M_\odot$. The maximum velocity at
the front of the ejecta is set to $20,000 \km \s^{-1}$. Its outer
radius from the center of explosion is set to almost touch the
He~WD.  {{{Ideally one should start from a real explosion of the CO WD. But we limit ourselves in the present study to explore the basic processes. We estimate the internal energy as follows. At shock breakout, about half the energy is thermal, half is kinetic.
As the gas expands, thermal energy drops as $1/r$.
By the time it hits the He~WD the thermal energy is one third of its initial value. Most of it went to accelerate the gas to almost the terminal velocity. The kinetic energy is now $5/6$  of the initial energy, and the thermal energy is $1/6$.
Over all, the kinetic energy is
5 times or more higher than the thermal energy. In the simulations we therefore set the thermal energy to be $0.2$ of kinetic energy at $t=0$ from the start of the simulations.  
We also simulated cases where the initial temperature was set to a very low value, and found no significant differences from the results presented here (see version V1 of this paper on astro-ph). }}}
In the figures described below time is measured from the moment at which the CO explodes. 
{{{{We ran our simulations with two different chemical compositions. In a first set of simulations we assumed that the ejecta was entirely made of nickel, while in a second set we adopted C/O. We found that our results are not sensitive to the adopted composition. (see version V1 of this paper on astro-ph for figures with C/O composition).}
}}}
Finally, we mention that radiative
cooling and photon diffusion are not important for the problem
simulated here, and hence have not been
included in our calculations.

\subsection{A low-mass helium WD}
\label{subsec:entrain}

In the case in which a low-mass He-WD is considered, nuclear
reactions are not significant and  three distinct stages of the
interaction can be differentiated. (i) The early interaction
phase, when temperatures of the shocked gas are at maximum, and
the ejecta flows around the He~WD. (ii) The intermediate phase,
when the shock breaks out from the back of the WD and ejects
helium from it. (iii) The late time phase, when  expansion is
homologous until the ejecta sweep a non-negligible ambient mass
and adopts the shape of an old SNR. We ran the simulations using
both the low- and high-resolutions grids. This was done for
checking numerical convergence. As mentioned earlier, the low-resolution
grid was designed to cover a larger region around the interacting
WDs, and thus was used to follow the evolution of the SNR  at late
times. In the overlapping regions,  the results of the two
simulations with different resolutions  were found to be the same.

\begin{figure}
\begin{center}
\includegraphics[width=1.0\textwidth]{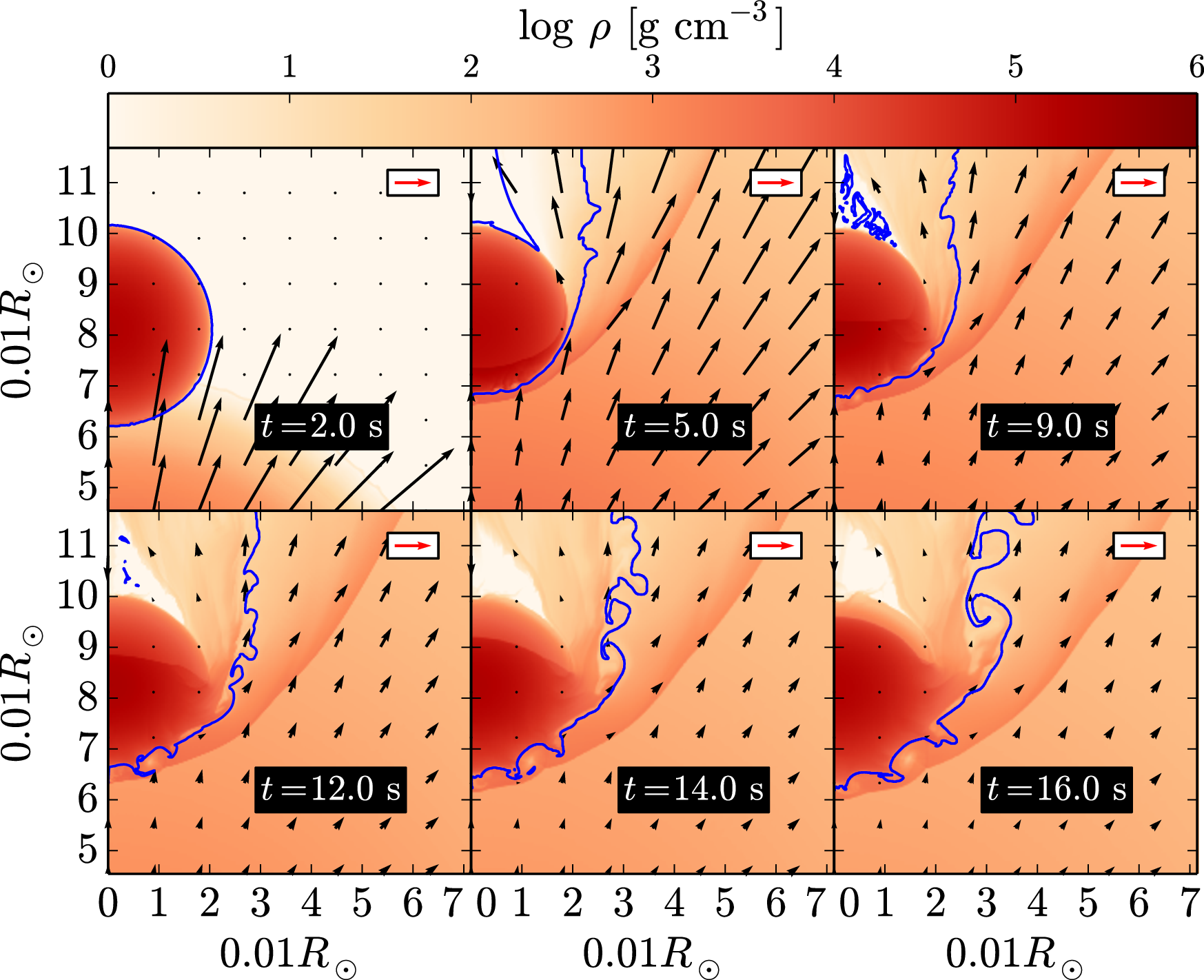}
\caption{Density maps in the meridional plane  at 6 times for the
case in which a $0.2 M_\odot$ WD is adopted. The time elapsed
since explosion is indicated in each panel. The simulation starts
$2 \s$ after explosion. The symmetry axis is along the left edge,
and the origin of the grid (outside the plots) is at the center of the exploding
CO~WD. The blue line encloses the volume where the local helium
mass fraction is $Y>0.5$; this represents the He~WD and the
material removed from the He~WD. Prominent features include a
shock running around the WD, and the formation of a dense conical
surface in the expanding ejecta. The shock just reaches the back
edge of the He~WD at $t=16 \s$. Temperatures and densities are too
low to drive any significant nuclear burning. The plots are from
the high-resolution run. The lower resolution simulation results
in a similar structure. Velocity is proportional to the arrow
length, with the inset showing an arrow for $10,000 \km \s^{-1}.$
{{{  Note the very fast gas at the outskirts, having velocities larger than the initial speed of $20,000 \km \s^{-1}$. This very low mass gas was accelerated by the initial thermal energy that was non-negligible. When the ejecta is inserted with low temperatures no such velocities are achieved; the differences from the present run are very small (see version V1 on astro-ph). }}}
}
  \label{fig:DensP1LowMass}
\end{center}
\end{figure}

{\it The early stage.} In Fig. \ref {fig:DensP1LowMass} we present
the density and velocity maps at several times from $t=2 \s$ (2
seconds after explosion) to the time instant at which the shock
that runs through the He~WD reaches the backside of the He~WD
($t=16 \s$). The SN ejecta hits the WD and flows around it,
forming a dense surface with a 3D conical shape. { {{{ Such
dense conical surfaces appear in the 2D simulations of
\cite{Panetal2010} and of \cite{Panetal2012a} where non-degenerate
companion stars were used.
}}}} In our 2D grid the dense shell has a shape of two dense
stripes on the meridional plane, one at each side of the symmetry
axis. Note that as mentioned in section \ref{sec:ignition} the
temperatures and densities are too low to drive any significant
nuclear burning.

\begin{figure}[h!]
\begin{center}
\includegraphics[width=1.0\textwidth]{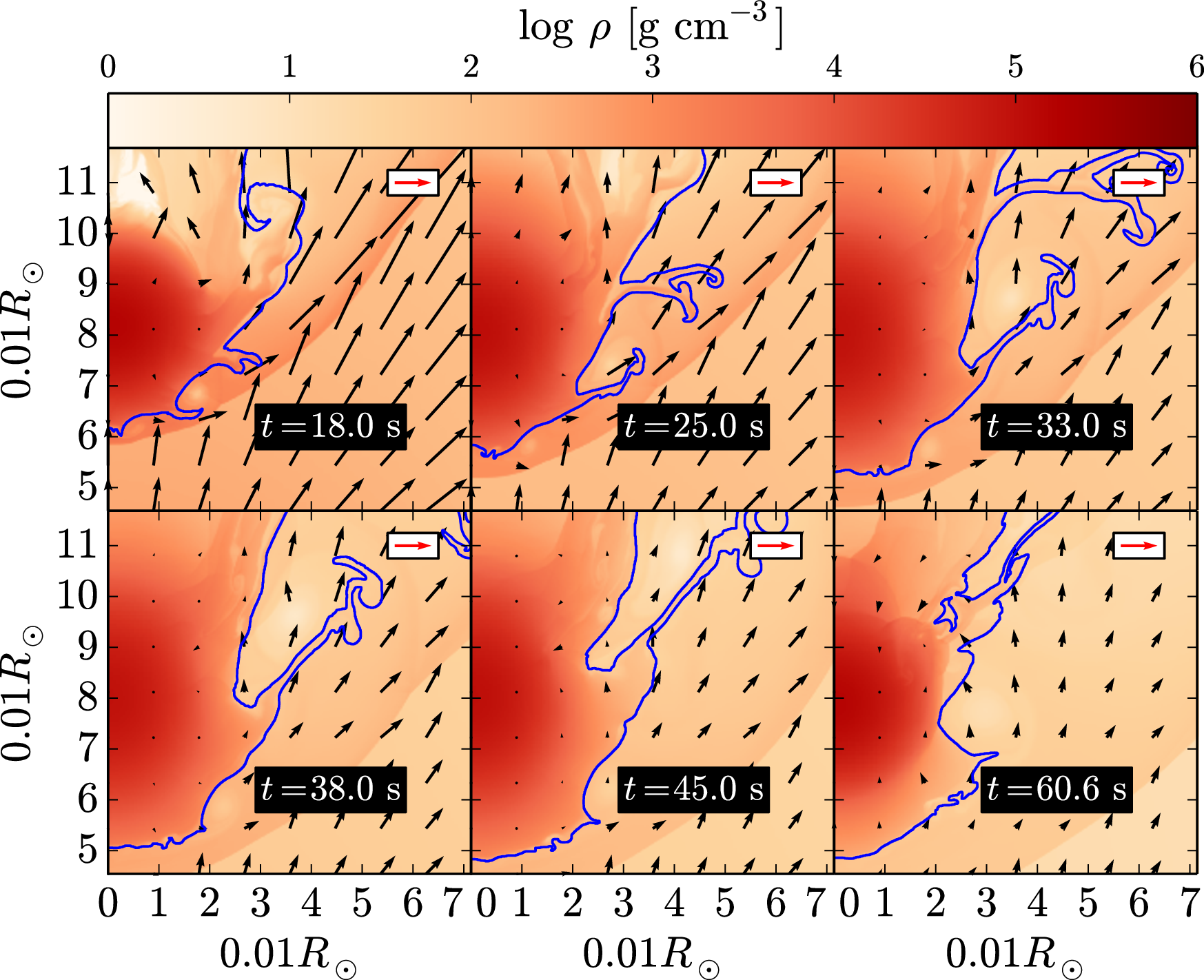}
\caption{Same as Fig. \ref{fig:DensP1LowMass} for later times, the
intermediate stage. The shock breaks out from the rear of the WD,
ejecting helium. Only $0.003 M_\odot$ of helium escapes while most
of the helium falls back on the WD as can be seen in the last
panel. The plots are from the low-resolution run. Velocity is
proportional to the arrow length, with the inset showing an arrow
for $10,000 \km \s^{-1}.$}
  \label{fig:DensP2LowMass}
\end{center}
\end{figure}

{\it The intermediate stage.} In Fig. \ref{fig:DensP2LowMass} we
show the flow after the break-out of the shock from the back side
(down flow) of the He~WD, and the consequential helium outflow.
Most of the ejected helium falls back to the WD as can be seen in
the last panel. Only $0.003 M_\odot$ of helium escapes and flows
outward near the symmetry axis, too small to be observed with
current observational means. The strong concentration at the axis
is a numerical effect. The volume inside the dense conical shell
is a region of low density ejecta. The dense conical surface
continues to expand and more or less preserves its shape in
homologous expansion. The homologous expansion continues until the
interaction with the ambient gas -- the interstellar medium (ISM)
or a circumstellar matter (CSM) -- starts to shape the outskirts
of the ejecta.

{\it The late stage.} We are interested in the morphology of the
ejecta at hundreds of years after explosion. For numerical
reasons, we let the ejecta interact with an ambient medium close
to the explosion site. As the ejecta expansion is already
homologous with high Mach numbers ($\ga 10$) at the end of the
intermediate stage, the morphology obtained here at the late stage
and on a scale of several solar radii represents quite well the
expected morphology hundreds of year later and with a much larger
size (a few pc). For the scaled numerical study of the
ejecta-ambient gas interaction we set the ambient density to be
$0.01 \g \cm^{-3}$, and follow the expansion until $t=492 \s$,
when the medium mass intercepted by the ejecta is $\sim 1
M_\odot$. The interaction of the dense conical surface with the
ambient gas forms a circle of high pressure, with its center on
the symmetry axis (half of this circle is into, and half out of,
the page). This high pressure circle accelerates  gas, both ambient
and ejecta, toward the relatively empty cone (toward the symmetry
axis). This gas and the helium along the symmetry axis, determine
the flow structure within the cone.
\begin{figure}[h!]
\begin{center}
\includegraphics[width=1.0\textwidth]{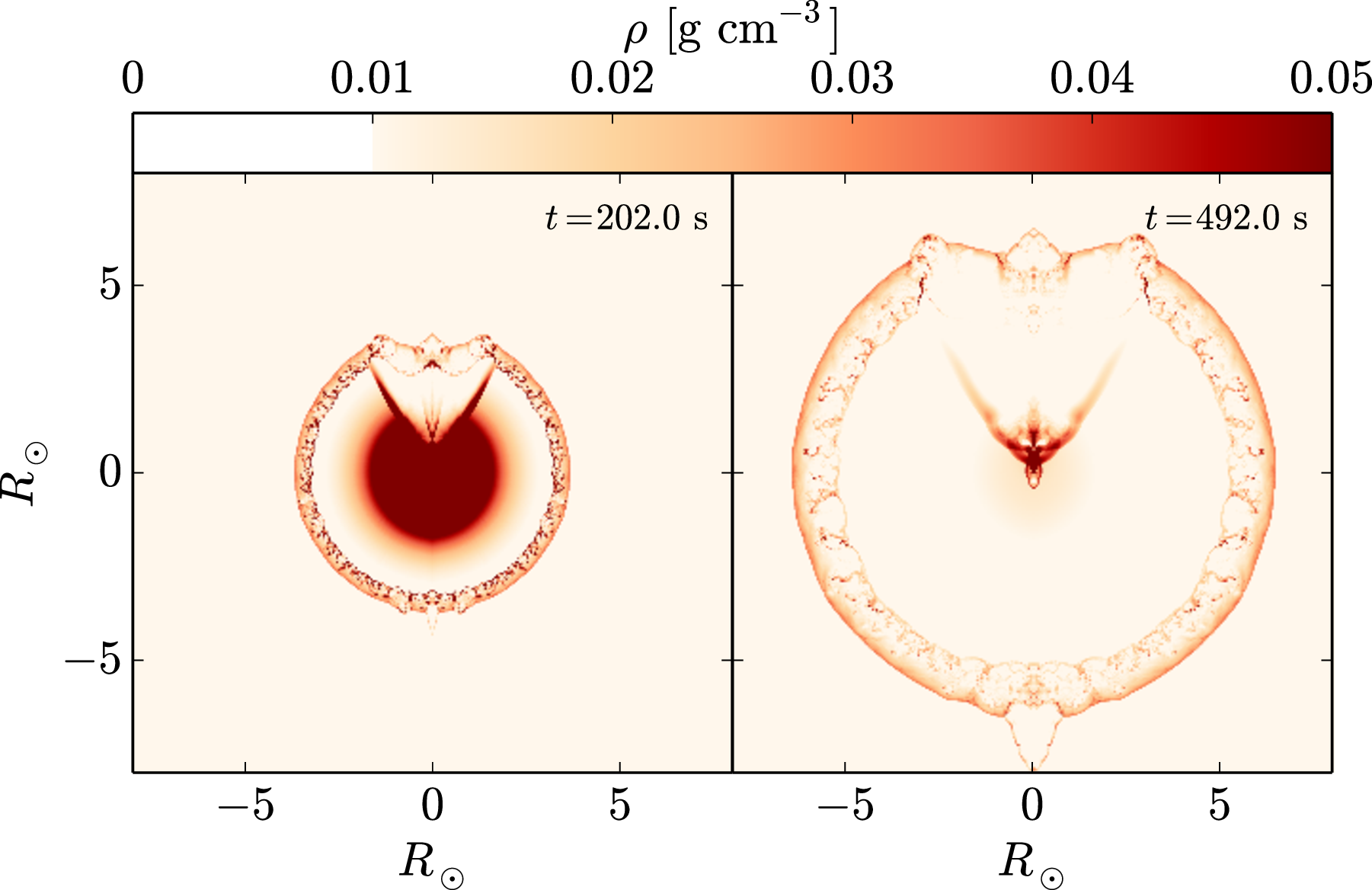}
\caption{Density maps in the meridional plane at 2 late times for
the case in which a $0.2 M_\odot$  WD is adopted. The
computational grid was folded around the axis to present the
entire meridional plane. A homologous expansion of the ejecta,
with a Mach number $>10$, has developed by the beginning of this
evolutionary phase, with a dense conical surface surrounding a
conical volume almost completely devoid of SN ejecta. The ambient
gas density is fixed by our requirement that at the end of the
simulation the ejecta sweeps a substantial mass (see text). At the
end of our simulations, $t=492 \s$,
the SN ejecta has swept $1 M_\odot$ of ambient gas. As the outflow
of the ejecta is already homologous, the morphology obtained here
mimics that at hundreds of years later. The small features along
the symmetry axis itself, both at the top and bottom of the SN-ISM
interaction, are numerical artifacts.}
  \label{fig:DensP3LowMass}
\end{center}
\end{figure}

{{{{ The morphological changes due to this flow depend on
the swept ISM mass in front of the dense conical surface, $M_s$. The
dissipated energy when the swept ISM mass is lower than the ejecta
mass that interacts with it $M_s<M_e$, is approximately $E_d \simeq
0.5 M_s v^2$, where $v$ is the radial speed of the ejecta. If a
fraction $\eta$ of this energy goes into azimuthal (tangential)
motion, then the azimuthal speed $v_\theta$ is given by $0.5 M_e
v^2_\theta \simeq \eta E_d$, from which we find $v_\theta \sim \eta
(M_s/M_e)^{1/2} \eta^{1/2}$. This is a crude expression, which nonetheless  shows that the filling of the empty cone depends mainly
on the total swept ISM mass, and not on the ISM density which is higher in our simulation due to numerical limitations. }}}}

To form a synthetic map (in radio, X-ray synchrotron, or thermal
X-ray), we integrate over density squared along the lines of
sight, but considering only shocked, hot gas,
 \begin{equation}
 I(x,y) \equiv \int [\rho(x,y,z)]^2 dz,
 \label{eq:I1}
 \end{equation}
where $x,y$ are the coordinates on the plane of the sky and $z$ is
taken along line of sight. {{{{The interaction regions are where
synchrotron emission will be formed. Although here the plots are
given shortly after explosion, in this paper we mimic the structure
hundreds of years after explosion, when radioactive decay is very
small and does not play a role in forming the hot regions. }}}}

\begin{figure}[h!]
\begin{center}
\includegraphics[width=0.45\textwidth]{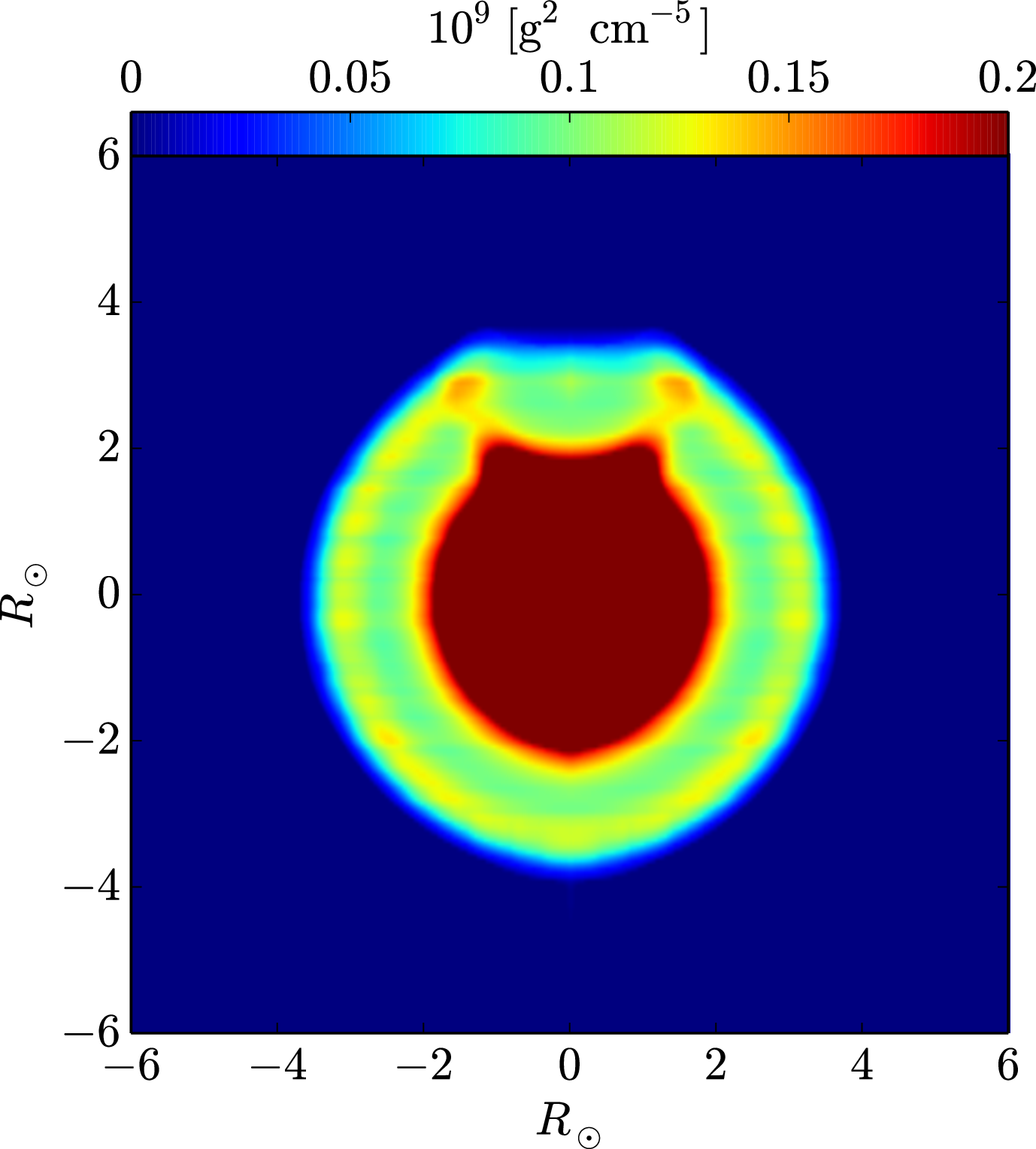}
\includegraphics[width=0.45\textwidth]{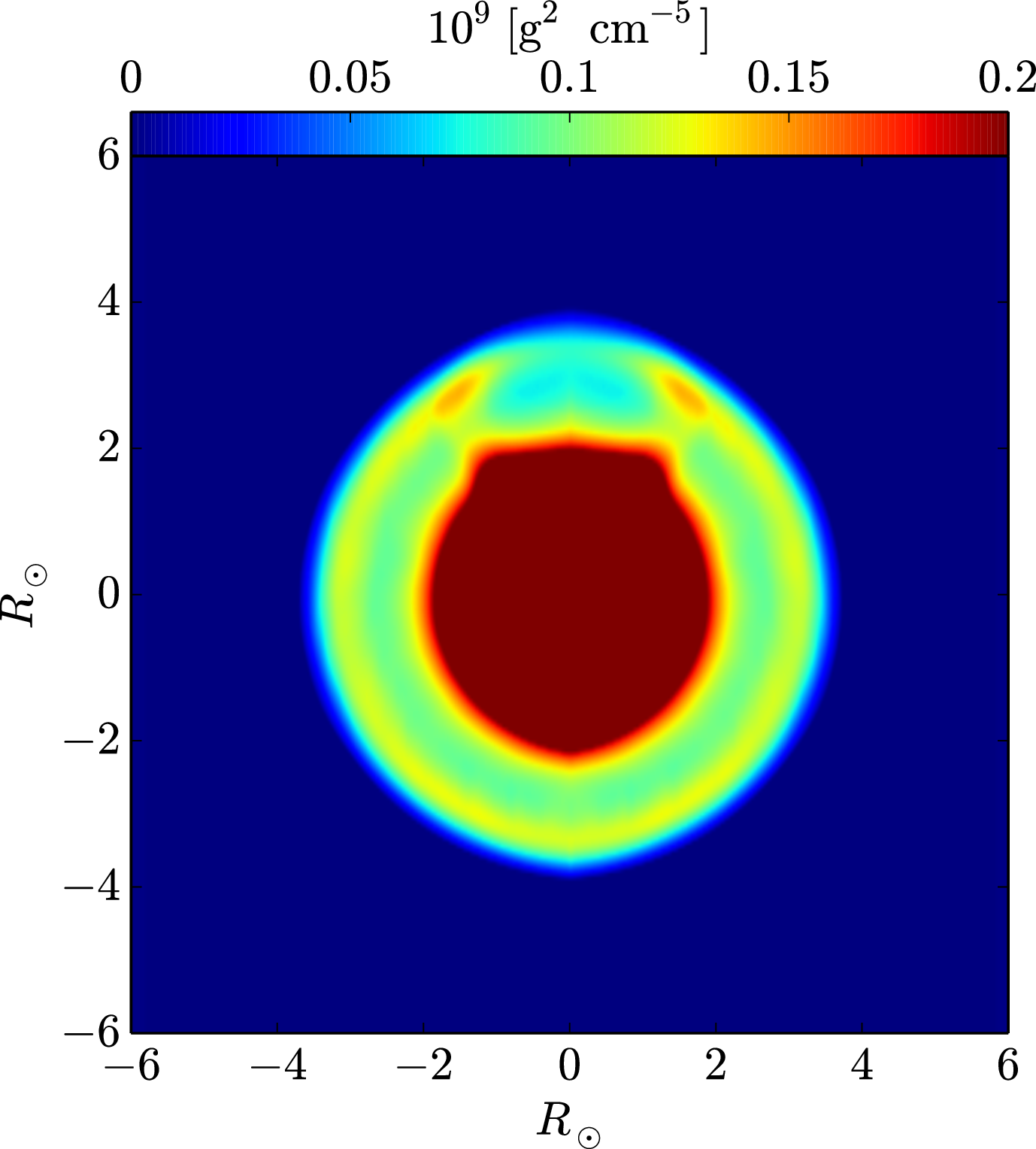}
\includegraphics[width=0.45\textwidth]{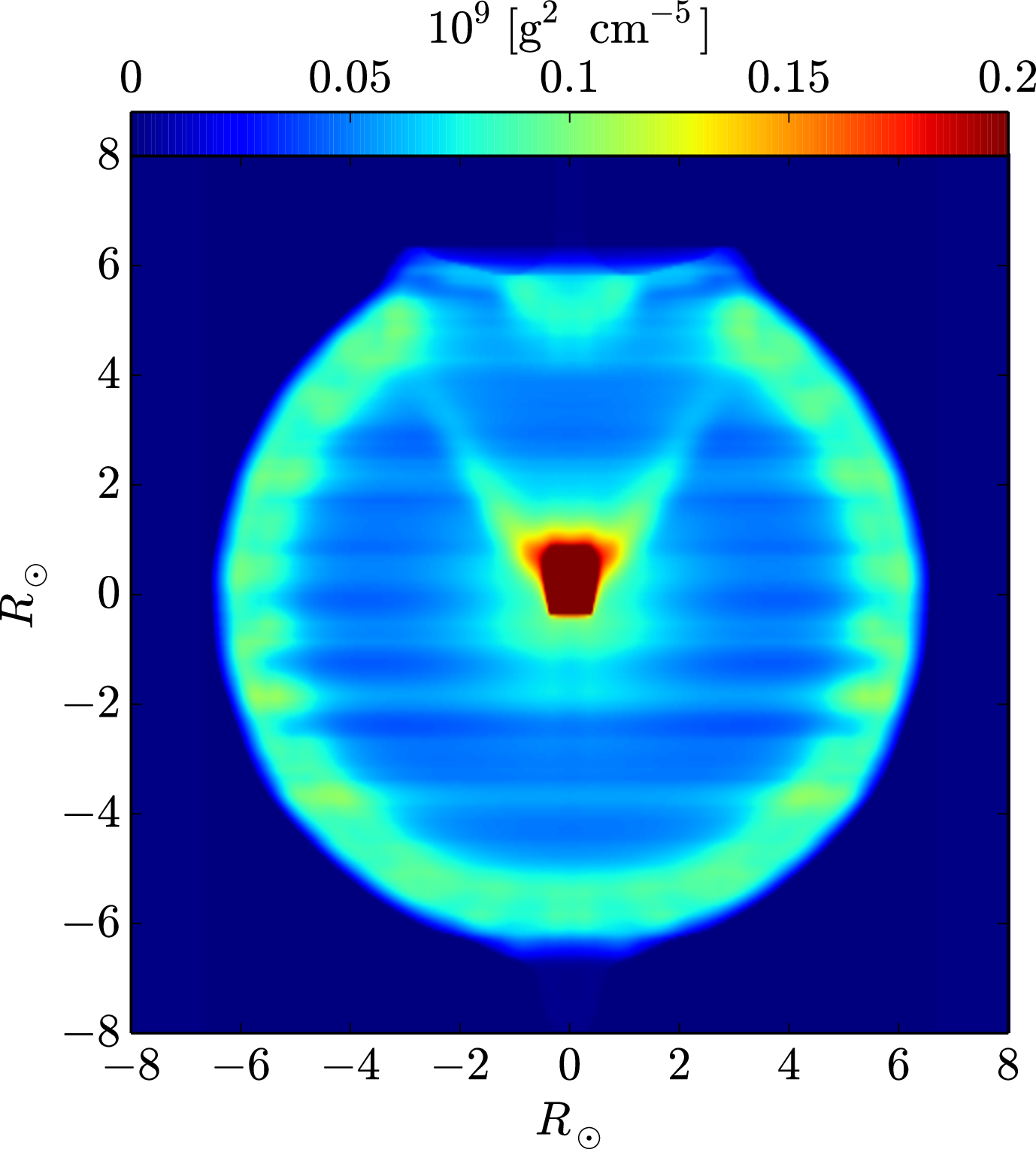}
\includegraphics[width=0.45\textwidth]{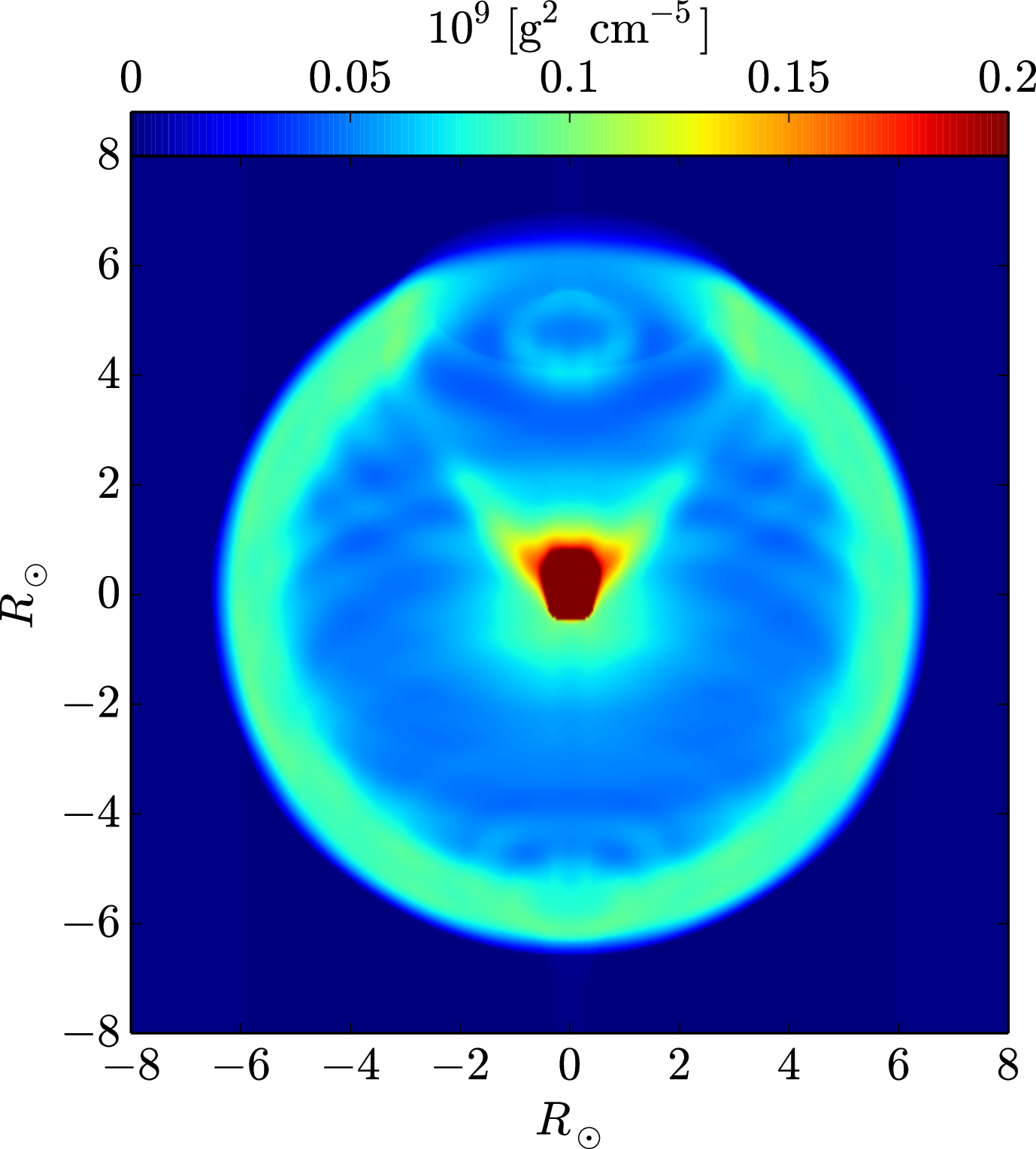}
\caption{Synthetic observed morphology (eq. \ref{eq:I1}) of the
resulting SNR for the case of a low-mass He~WD.  We show the
intensity map described in the main text, and only
  for the high-temperature gas. The   $x$ and $y$  coordinates are on the  plane of the sky,   and the $z$ coordinate is taken  along line of
  sight. Two inclinations are presented,
the symmetry axis is in the plane of the sky (left), or at
$30^\circ$ to the plane of the sky (right). These are presented at
two times, namely when the swept-up ambient masses are $\sim 0.1 M_\odot$
($t= 202 \s$ upper panels), and $\sim 1 M_\odot$ ($t=492 \s$ lower
panels).
    As the outflow of the ejecta is already homologous at the beginning of this phase,
the morphologies obtained here mimic that at hundreds of years
later when the ejecta interacted with $\sim 0.1-1 M_\odot$ of
homogeneous ambient medium (CSM or ISM).
 }
  \label{fig:ImageP3LowMass}
\end{center}
\end{figure}

\begin{figure}[h!]
\begin{center}
\includegraphics[width=0.5\textwidth]{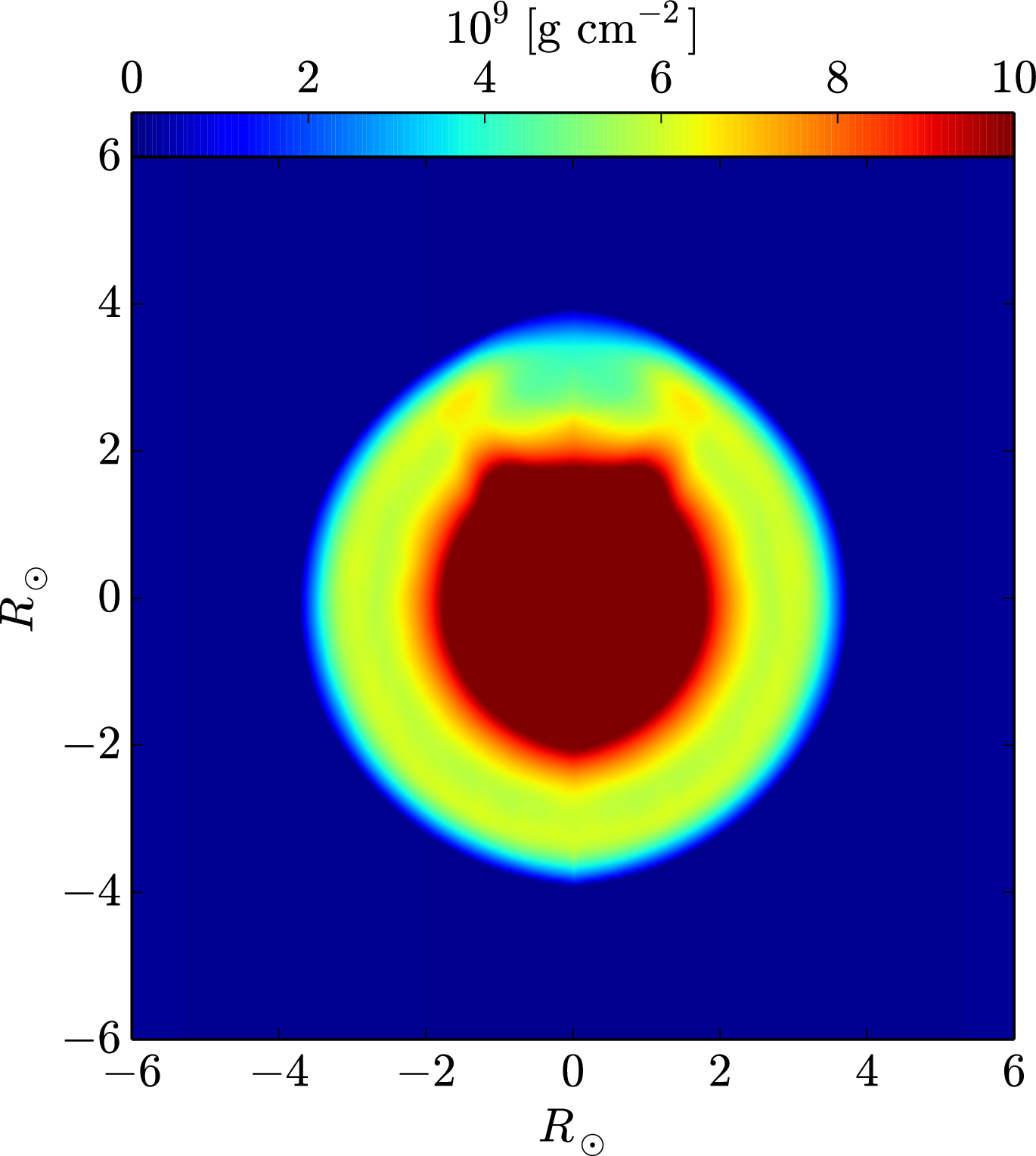}
\includegraphics[width=0.5\textwidth]{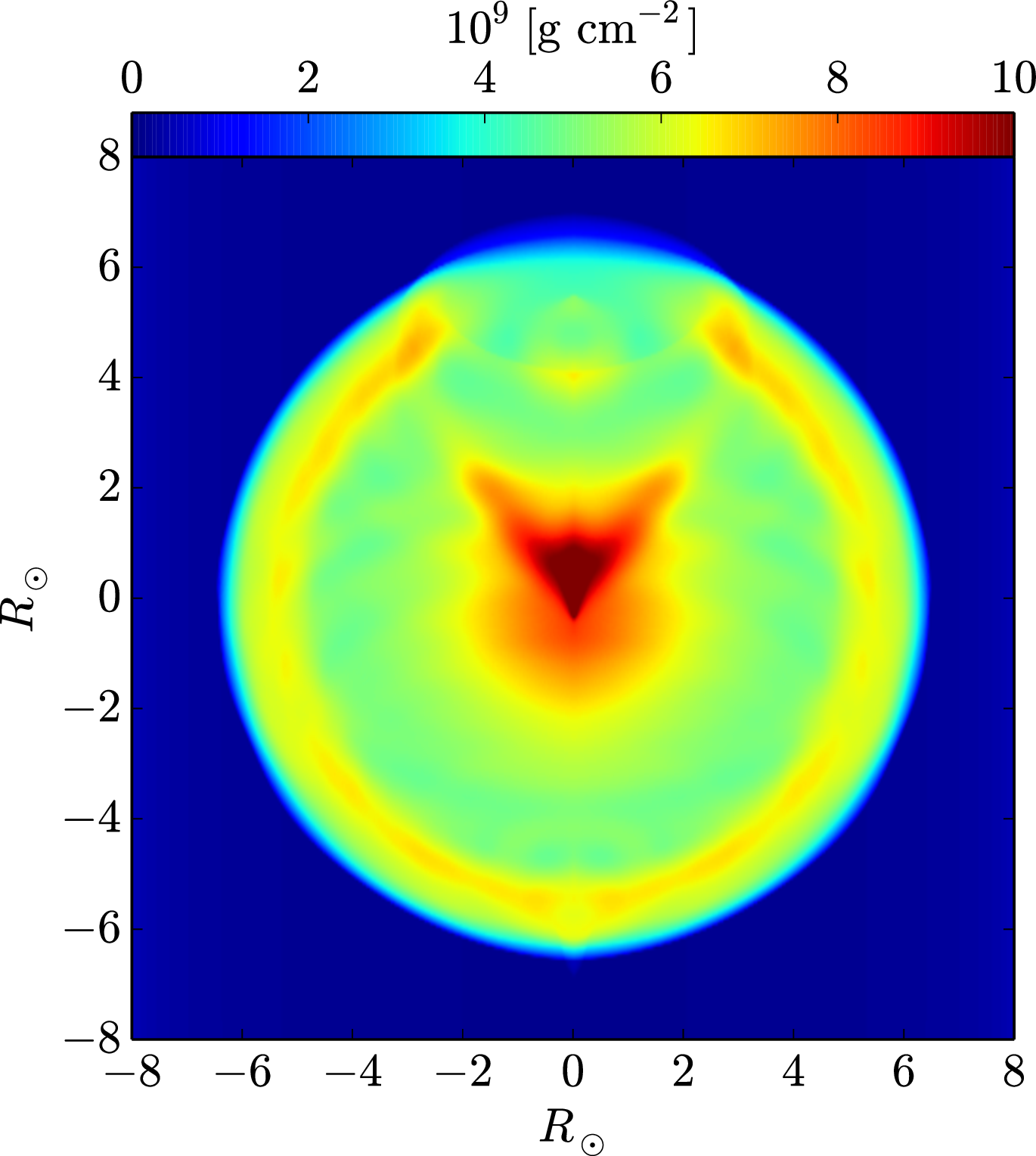}
\caption{The integrated ejected mass (eq. \ref{eq:I2}) for the two
times as in Fig. \ref{fig:ImageP3LowMass}, and for the symmetry
axis  at $30^\circ$ to the plane of the sky.
{{{ Note the very low fraction of ejecta in the shadow behind the He WD (upper part in the figures) close to the edge of the remnant. }}} 
}
  \label{fig:ImageP3OnlyEjectaLowMass}
\end{center}
\end{figure}

The obtained `intensity maps' are presented in Fig.
\ref{fig:ImageP3LowMass}. Two inclinations are presented, the
symmetry axis is in the plane of the sky (left), or at $30^\circ$
to the plane of the sky (right). These are presented at two times
when the swept-up ambient masses are $\sim 0.1 M_\odot$ ($t= 202
\s$ upper panels), and $\sim 1 M_\odot$ ($t=492 \s$ lower panels).
In Fig. \ref{fig:ImageP3OnlyEjectaLowMass} we present the integral
of the density but only for the ejected mass,
 \begin{equation}
 N_{\rm eject}(x,y) \equiv \int [\rho(x,y,z)_{\rm eject}] dz
 \label{eq:I2}
 \end{equation}

The prominent features of the  SNR  when the symmetry axis is
close to the plane of the sky before the swept ISM gas is too
large are the following ones.
 (a) A `flat front' of the conical region (upper part in the figure which
is the initial direction of the He~WD); (b) A region of lower
intensity at that flat front relative to the rest of the SNR
front; (c) A dense conical surface in the interior; (d) The inner
volume of the conical surface is almost completely devoid of
ejecta gas. The first two features  fade as more ambient gas (ISM)
is swept. Let us note that the main result here does not depend
much on whether the He~WD is younger and hotter, hence has a
larger radius. It will simply have somewhat larger orbital
separation. But as the double-detonation model requires stable
Roche lobe overflow (RLOF), the solid angle covered by the He~WD
will be about the same, and so is the conical shape formed behind
it {{{{ (see \citealt{Mariettaetal2000}). Here we find the
angular size (from symmetry axis to conical surface) of the cone
to be $\sim 35^\circ$. \cite{Mariettaetal2000}
found in their study of the single-degenerate scenario that the
companion creates a `hole' in the supernova debris with an angular
size of $\sim 30-40^\circ$, depending on the part of the ejecta,
and \cite{Pakmoretal2008} found an angular size of $\sim 23
^\circ$. Most similar to our structure of the cone are the results
of \cite{Panetal2010}, where the angular size of the cone is $\sim
40^\circ$, and of \cite{Panetal2012a} where in many cases the
angular size of the cone is $\sim 40^\circ$ (in some 3D
simulations there is no well defined cone). All these results agree
with each other within the range of different initial parameters.
}}}}

{{{{ Although some SNRs show some dipole deviations from
sphericity,  we are not aware of any SNR that shows such a conical imprint morphology.}}}} One might think of SN1006, but examining the prominent
features we find that SN1006 cannot be explained by such an
interaction. (a) SNR SN1006 has a flat front. However, there is a
hydrogen-rich optical filament along the flat front. The flat
front seems to have been formed by an asymmetrical external
interaction formed by asymmetrical ISM. (b) In SN1006 the X-ray
intensity of the flat front is lower than the front on the
orthogonal directions, but not lower than the other side of the
SNR (e.g., \citealt{Winkleretal2014}). Also, SN1006 does not show
a uniform intensity along the spherical parts not including the
flat front. (c) A dense conical surface in the interior is not
observed in SN1006 (e.g., \citealt{Winkleretal2014}) (d)  As can
be seen from figure 9 of \cite{Winkleretal2014}, the volume behind
the flat front is rich in neon and oxygen, and it is not poor in
ejecta. We conclude that the structure of the SNR SN1006, despite
the flat front on one side, is incompatible with the morphology
expected from the double-detonation scenario.

{{{{ The results of asymmetrical SNR obtained here applies
to all single-degenerate scenarios as well. The DD scenario also
leads to asymmetrical explosion if it occurs too shortly after the
merger of the two WDs. Overall, it seems that the symmetrical
structures of most SNRs Ia hint that when it explodes the WD is
all alone. This is compatible with the CD scenario. In cases where
a circumstellar gas is present and influences the SNR morphology,
e.g., in forming two opposite `ears' as in the Kepler SN remnant,
the CD scenario seems to do better than other scenarios as well
\citep{TsebrenkoSoker2015b}. }}}}

{{{{ In some SNRs one can identify two opposite `ears' that
divert the SNR from being spherical (see
\citealt{TsebrenkoSoker2015b} for a list of objects). These `ears'
might be formed by jets in the pre-explosion evolution, as
expected for some SNRs in the CD scenario
\citep{TsebrenkoSoker2015b}. These SNRs are not perfectly
spherical, but the asymmetry is like a quadrupole, and not as a dipole as
expected if a companion influence the shaping of the SNR. }}}}

{{{{ A word of caution is in place here. Our conclusions
hold as long as there are no processes that erase the asymmetry
caused by the companion. If the initial asymmetry is large, e.g.,
as proposed by \cite{Maedaetal2010}, then the morphology of SNRs
discussed above implies that there is a process that erases
asymmetry. For example, radioactive heating of dense regions can
cause them to expand and fill empty regions. However, three points
should be made regarding the homogenizing effect on the flow by
radioactive heating. (1) The change in velocity and density cause
a deviation from the purely homologous density profile of about 10
per cent \citep{PintoEastman2000, Woosleyetal2007,
Noebaueretal2012}. Such small variations will not erase the dipole
asymmetry. (2) The nickel is concentrated in the center, while we
are interested in the outer layers that are first to interact with
the ISM. (3) The observed very low level continuum polarization at
the first few weeks in SN~2012fr points to a symmetrical explosion
that is inconsistent with the merger-induced explosion scenario
\citep{Maundetal2013}. Namely, it seems that explosion is not far
from spherical from the beginning. }}}}

{{{{ Over all, despite the caution one must take at this
stage, the assumptions and approximations made here lead to a fair
representation of the SNR that result from the double detonation
scenario with low mass He~WD as the donor.  }}} }

\subsection{A massive helium WD}
\label{subsec:MassiveWD}

In this  case we place  a $0.43  M_\odot$ He~WD at  closer
distances than   the    $0.2   M_\odot$    one,   as   described
in   section \ref{subsec:setup}.  We  find that  the helium WD  is
ignited when the distance of  its center to the center of the CO WD is $\la  3.1 \times 10^9 \cm = 0.045 R_\odot$, and that
practically no burning occurs if it is placed at larger distances.

\begin{figure}[h!]
\begin{center}
\includegraphics[width=1.0\textwidth]{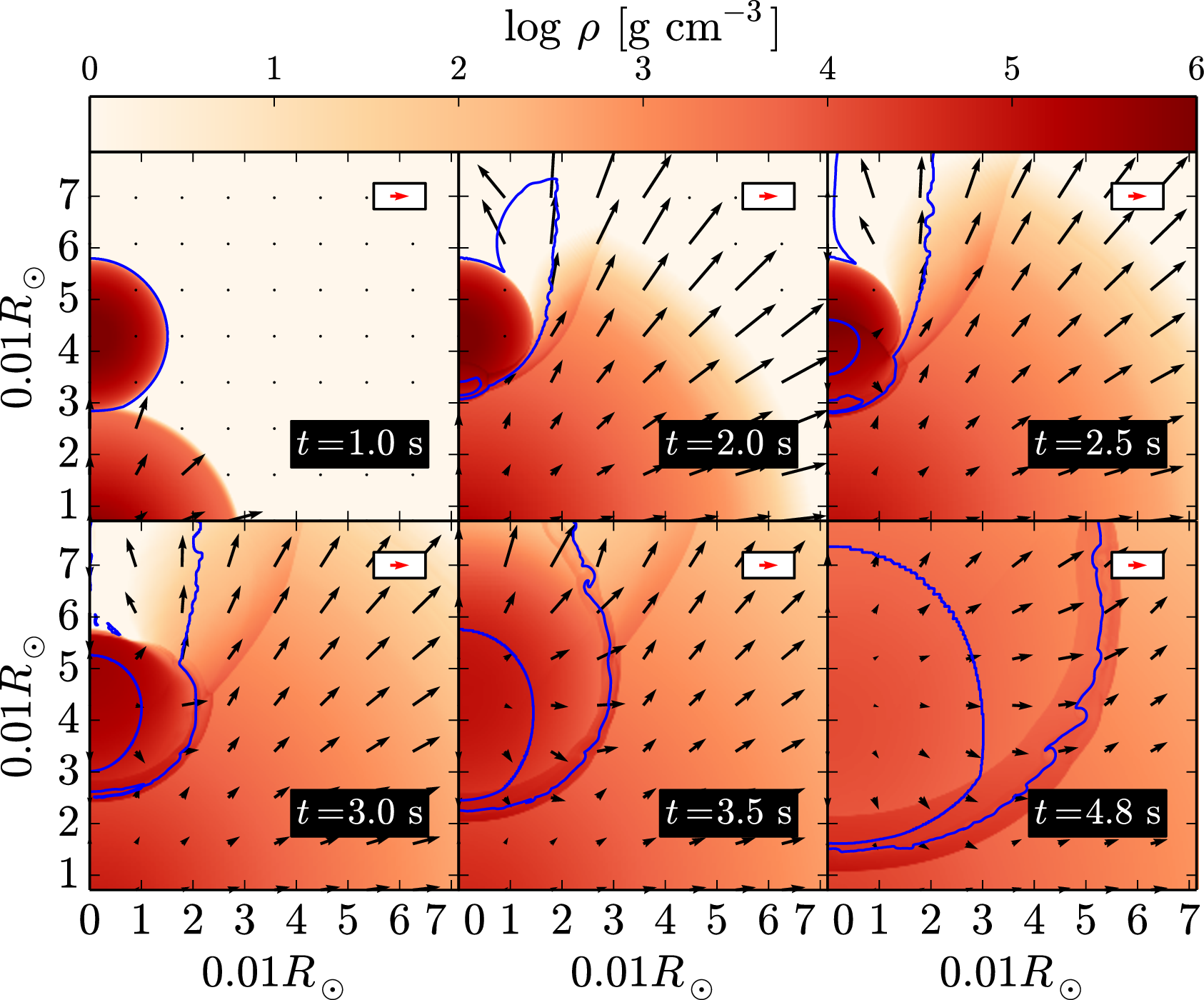}
\caption{Density maps in the meridional plane at six times for a
He~WD of $0.43 M_\odot$ at an initial distance of its center to
the center of explosion of $0.045 R_\odot$. Note that at $t=2 \s$
helium is ignited and  an explosion occurs in the He~WD. The
velocities are proportional to the arrow length, with the inset
showing an arrow for $10,000 \km \s^{-1}.$ 
}
  \label{fig:ImageDensity3e9HighMass}
\end{center}
\end{figure}

\begin{figure}[h!]
\begin{center}
\includegraphics[width=1.0\textwidth]{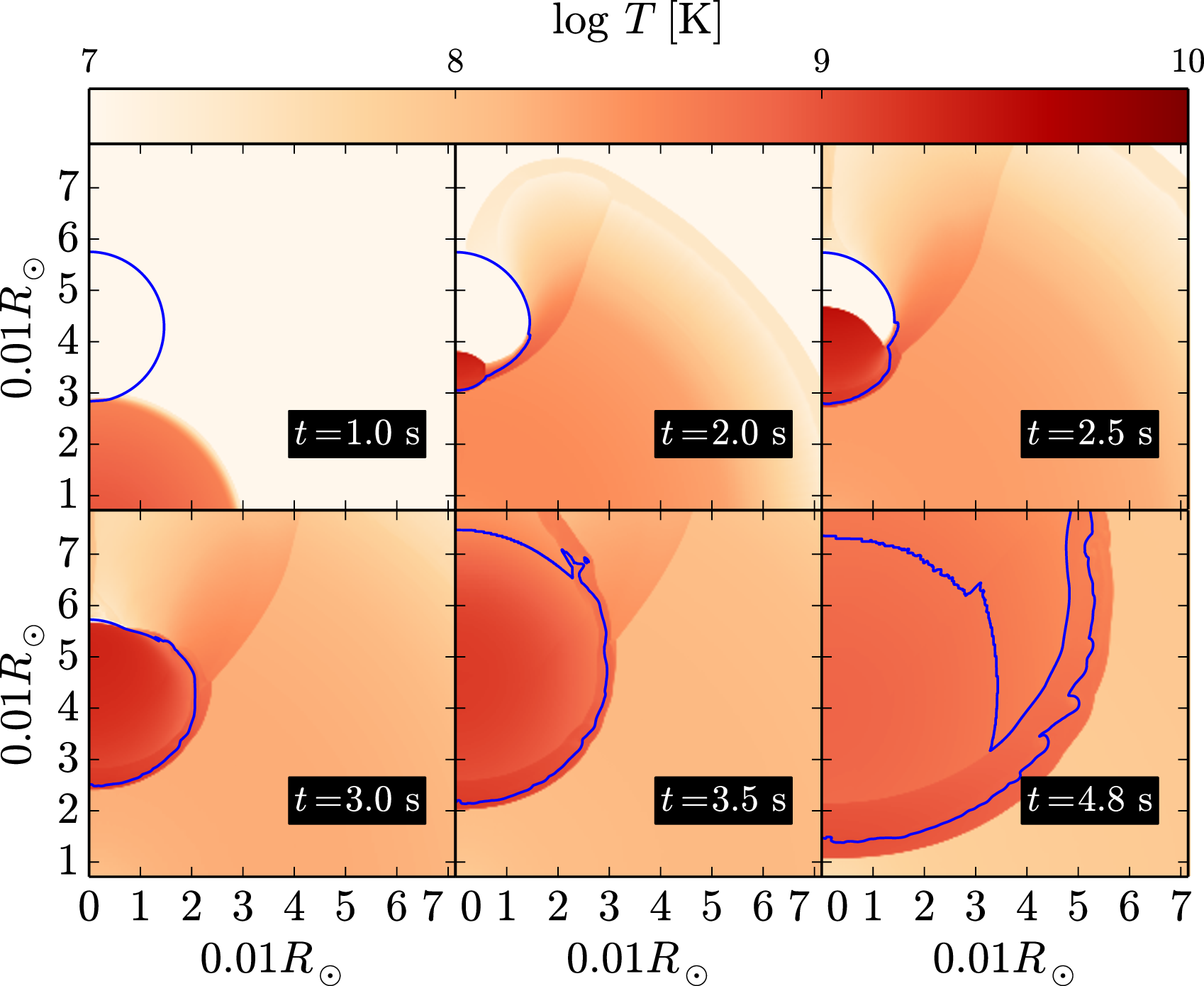}
\caption{Same as Fig. \ref{fig:ImageDensity3e9HighMass} but for
temperature.}
  \label{fig:ImageTemp3e9HighMass}
\end{center}
\end{figure}

\begin{figure}[h!]
\begin{center}
\includegraphics[width=1.0\textwidth]{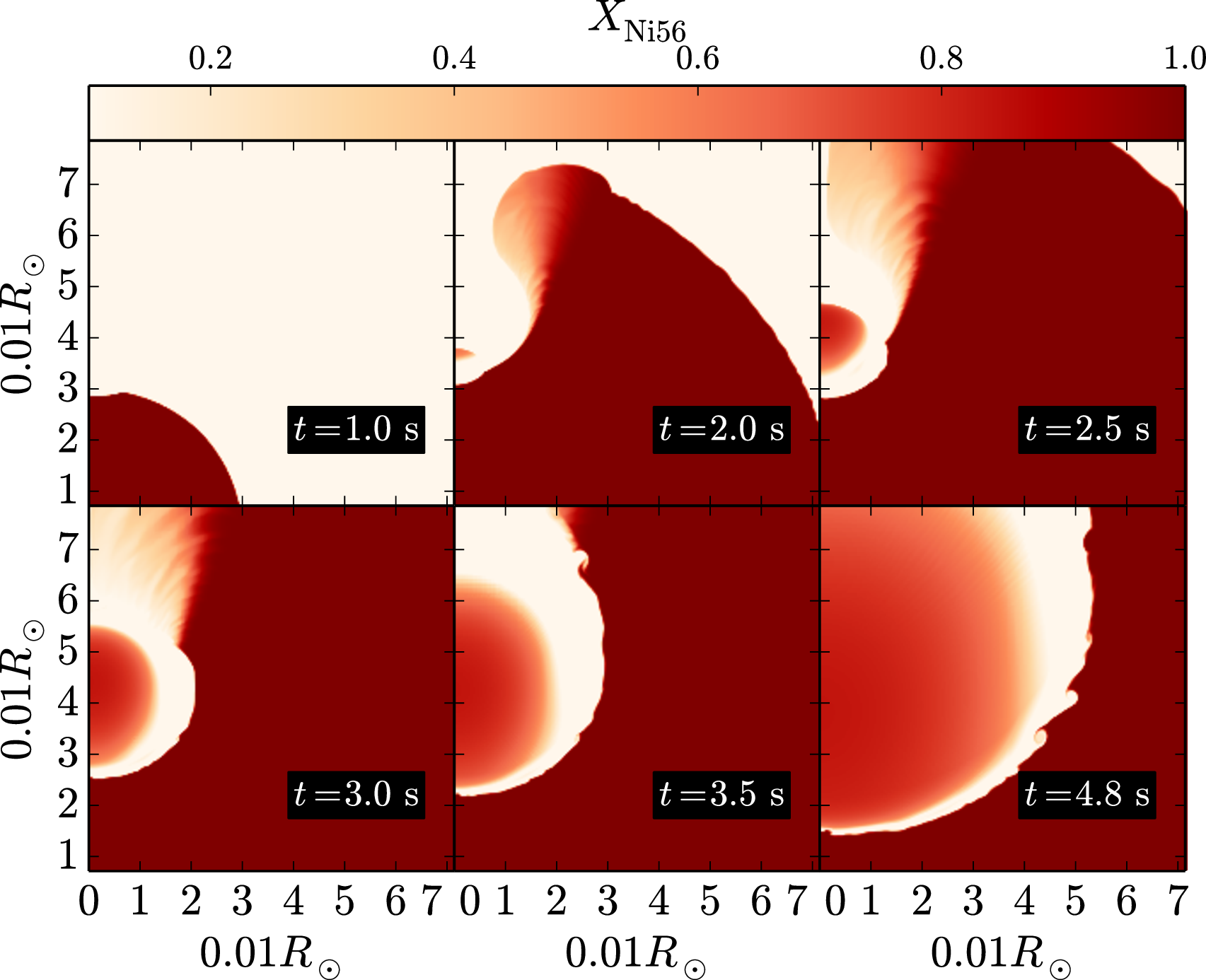}
\caption{Same as Fig. \ref{fig:ImageDensity3e9HighMass} but for the
nickel mass fraction. Ignition of helium in the He~WD occurs just
before $t=2 \s$.
{{{ The deep-red indicates the ejecta gas, that we took to be composed entirely of nickel. (Using CO composition for the ejecta does not change the results; see version V1 of the paper on astro-ph.). The lighter-red is the nickel mass fraction that is synthesized in the He WD. White regions are composed of He~WD gas that did not form nickel. }}} 
}
  \label{fig:ImageNicke3e9HighMass}
\end{center}
\end{figure}

 In Figs. \ref{fig:ImageDensity3e9HighMass} to \ref{fig:ImageNicke3e9HighMass}
we present the evolution of density, temperature, and nickel mass
fraction, of the ignited He WD at 6 different times, as indicated.
The initial distance of the center of the He WD from the center of
explosion is $0.043 R_\odot$.


\begin{figure}[h!]
\begin{center}
\includegraphics[width=1.0\textwidth]{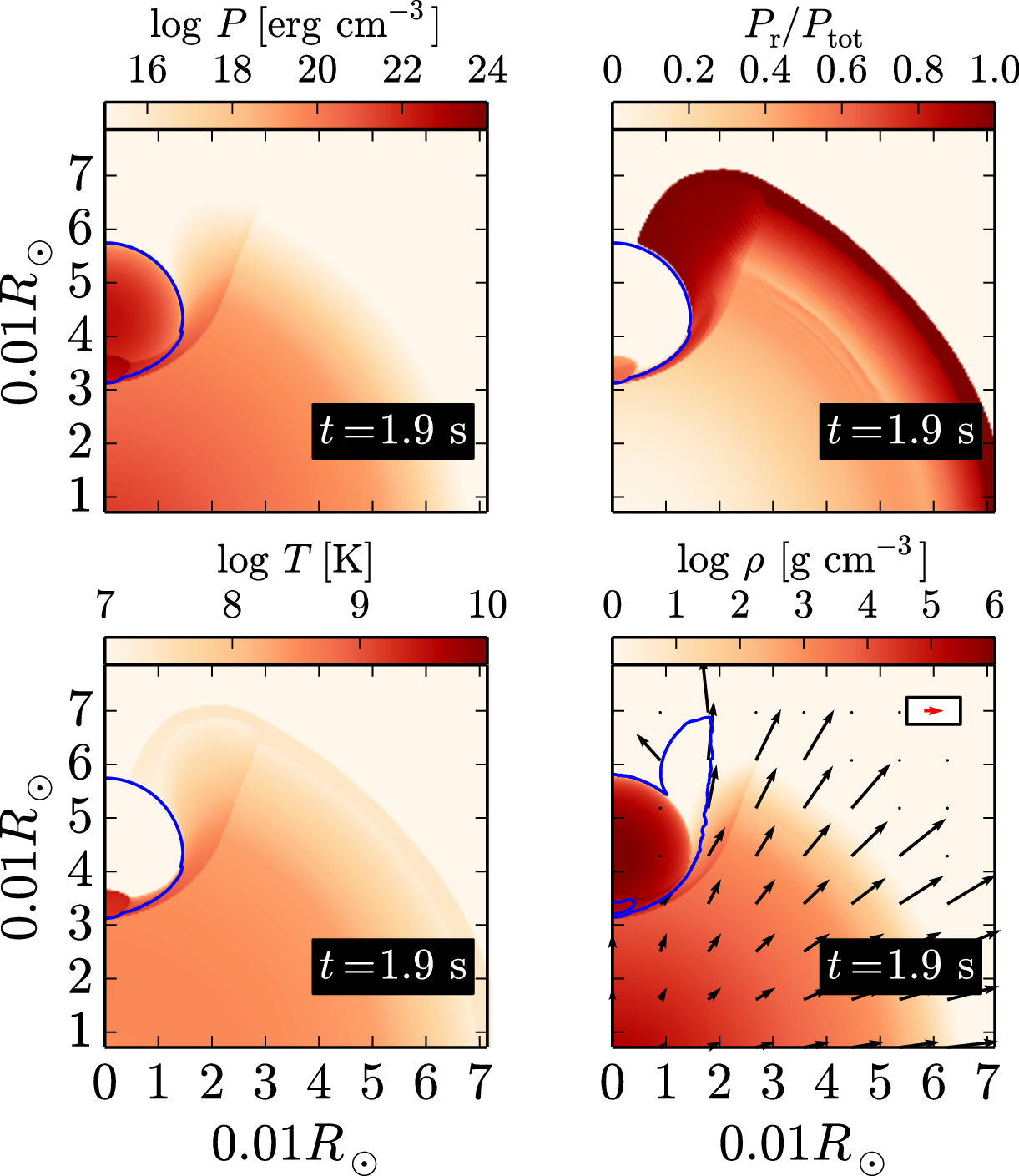}
\caption{Total pressure (top left), ratio of radiation to total
pressure (top right), temperature (bottom left), and density
(bottom right) at $t=1.8 \s$, just after He ignition. The velocities are proportional to the
arrow length, with the
 inset
showing an arrow for $10,000 \km \s^{-1}.$ The blue line shows
when the helium fraction is $Y=0.5$.
 The figure is for the case in which a He~WD of mass $0.43 M_\odot$ is adopted.}
  \label{fig:pressure043}
\end{center}
\end{figure}

Note that this calculation shares some features in common with the
evolution in the case in which a low-mass He~WD is considered, but
also some noticeable differences. In particular, although the
evolution of the hydrodynamical flow is apparently similar, the
key difference is the much larger temperatures attained during the
interaction between the ejecta and the He~WD. Ignition of helium
occurs just before $t=2 \s$, as can be seen in the lower panels of
Fig. \ref{fig:pressure043}. The ignited helium raises the
temperature and a thermonuclear detonation occurs, in accordance
with  the theoretical estimates presented in  section
\ref{sec:ignition}. By the last panel the explosion has ended.

It is interesting to note as well the important role of radiation
pressure in this simulation, as it should be expected given the
considerations explained in section~\ref{sec:ignition}. To
corroborate this, in the upper panels of Fig.
\ref{fig:pressure043} we show the total pressure (top) and ratio
of radiation to total pressure (bottom) at the time of helium
ignition, $t=2 \s$. It can be seen that at the ignition point the
radiation pressure dominates, but thermal pressure is not
negligible. Also, the total pressure in the ignition region is
$\sim 10^{22} \erg \cm^{-3}$, comparable to the estimate given in
equation (\ref{eq:pramm}) if we adopt $r_e=0.04 R_\odot$.

Given that the temperatures attained during
the interaction between the ejecta and the massive He~WD are
rather high, extensive nuclear processing occurs, and a
substantial amount of nickel is synthesized. Nickel first appears
in a region laying between the center of the He~WD and the surface
facing the ejecta. Note that after a few seconds most of the
material of the He~WD has been processed to nickel. This
contradicts observations, as the SNR will be highly asymmetrical,
as in the violent merger simulation presented by
\cite{Pakmor2012}. We find that not all helium is burned and $\sim
0.15 M_\odot$ of helium is ejected from the exploding He~WD. 
{{{ {This also contradicts some observations, e.g., \cite{MazzaliLucy1998} found a limit of $<0.1 M_\odot$ of helium in SN Ia 1994D.} }}}  {{{ { In a recent study \cite{Lundqvistetal2015} put a much stronger limit of $\la 0.01 M_\odot$ of ablated mass from a helium-rich companion to SN~2011fe and to SN~2014J. }}}}
 We conclude that the presence of a relatively close by, $a_0 \la 0.45
R_\odot$, He~WD donor to the exploding CO~WD leads to an explosion
that  has characteristics contradicting observations of SNe Ia.
Accordingly, the double-detonation scenario seems to do not apply to
normal SNe Ia.

{{{{ We have actually simulated here a `triple detonation
scenario'. The three stages are: He detonation on the surface of a
WD, then a CO detonation, and finally a He detonation in the He WD
companion. The outcome is a total ejected mass of about the
Chandrasekhar mass, although the two WDs were each much below the
Chandrasekhar mass. The ejected mass and synthesized nickel are
larger than those inferred for SN~2005E \citep{Peretsetal2010}, or
`calcium-rich gap' transients in general
(\citealt{Kasliwaletal2012}; for a recent list of transients see
\citealt{Perets2014}). We also expect iron group elements, which
are not generally observed in SN~2005E and the other
`calcium-rich gap' transients \citep{Peretsetal2010,
Kasliwaletal2012}. One of these gap transients have hydrogen
\citep{Kasliwaletal2012}, which is not expected in the
tripe-detonation scenario. Such transients are more likely to come
from helium detonation on a WD without ignition of the He~WD
companion  \citep{MengHan2014}.  }}}}

{{{{
The presence of helium might lead to classification of the event as a SN
Ib, but with high helium-burning products that will make it a peculiar SN Ib.    Such SNe might be related to the  peculiar low-luminosity SNe Ib with relatively strong Ca spectral lines (e.g., \citealt{Peretsetal2010, Foley2015}). 
\cite{Foley2015}, following  \cite{Peretsetal2010}, suggests that the progenitor system for these SNe is a double WD system where at least one WD has a significant He abundance. We here raise the possibility that some Ca-rich peculiar Ib SNe come from the triple-detonation scenario. This speculation deserves a separate study. In any case, we expect the triple-detonation scenario to be very rare. }}}}

\section{SUMMARY AND CONCLUSIONS}
\label{sec:conclusions}

We have studied the impact of the ejecta of an exploding CO WD on
the donor star in the double-detonation scenario for the formation
of Type Ia supernovae (SN Ia). We have done so for two masses of
the secondary He WD, namely $0.2 M_\odot$ and $0.43 M_\odot$,
assuming that the SN Ia ejecta is already in homologous expansion
when it hits the surface of the secondary WD. The first part of
our study was done using analytical estimates, while in the second
part of our work we performed full 2-dimensional hydrodynamical
calculations, employing the \texttt{FLASH} code. Our most relevant results
can be summarized as follows.

For the case in which a massive He~WD ($0.43 M_\odot$) is
considered, our analytical estimates predicted that the material
of the He~WD would  undergo a powerful thermonuclear runaway when
the ejected material of the exploding CO~WD interacts with outer
layers of the donor WD (Sect. \ref{sec:ignition}). Our analytical
predictions are confirmed by our detailed hydrodynamical
calculations that also give us the evolution with time of the flow,
where ignition occurs, the amount of nickel formed, and the mass
 of helium ejected by the interaction (Figs.
\ref{fig:ImageDensity3e9HighMass} - \ref{fig:pressure043}). In particular,
the mass of ejected helium ($0.15 M_\odot$) would have been easily detected in
observations, implying that this scenario seems to be ruled out for
standard SN Ia.

For the binary system containing a low-mass He~WD ($0.2 M_\odot$)
no significant nuclear processing occurs, and the evolution
consists of an almost pure hydrodynamical flow. The evolution can
be divided in three distinct phases. During the initial phase a
shock runs through the outer layers of the He~WD, and the SN
ejecta flows around the secondary star, forming a region with
conical shape (Fig. \ref{fig:DensP1LowMass}). In the intermediate
stage, just after the shock breaks-out from the back side of the
He~WD, some material from the He~WD is ejected but most of it
falls back at later times, while a conical dense surface continues
expanding (Fig. \ref{fig:DensP2LowMass}). Finally, during the late
stages of the evolution the SN ejecta interacts with the ambient medium, which we
numerically set to a very high density to mimic interaction with
the ISM hundreds of years later. During this phase the conical
flow previously described forms a ring of high pressure, which
accelerates material towards the low-density conical region (upper right panel of Fig.
\ref{fig:DensP3LowMass}).

The hydrodynamical evolution previously described has
observational consequences. In an attempt to model the morphology
of the resulting SNR we integrated the density squared of the hot
gas for two viewing angles and two times (Fig.
\ref{fig:ImageP3LowMass}. The integrated ejecta density is shown
in Fig. \ref{fig:ImageP3OnlyEjectaLowMass}). We found that the
shape of the SNR, that contains a prominent flat region in the
direction of the shadow of the He~WD, is at odds with known SNR
morphologies.

In conclusion, our study supports previous claims that the
double-detonation scenario can at best be responsible for a very
small fraction of all SN Ia. Specifically,
\cite{Piersantietal2013} claimed that the double-detonation
scenario can account for only a small fraction of all SN Ia,
because the parameter space leading to explosion is small.
\cite{Ruiteretal2014}, on the other hand, argued that the
double-detonation model can account for a large fraction of SN Ia.
For that to be the case, most ($>70 \%$) of the donors in the
study of \cite{Ruiteretal2014} are He~WD.  Our results show that
He~WD donors lead to explosions that are in contradiction with the
observed morphology of the SNRs of Type Ia SN, and that if the
He~WD is massive ($\sim 0.4 M_\odot$), not all helium is burned
and, consequently, would be spectroscopically observed, again in
contradiction with observations.

There is another severe problem with the double detonation
scenario \citep{TsebrenkoSoker2015b}. As \cite{Ruiteretal2014}
showed, most exploding WDs in the double-detonation scenario are
of mass $<1.1 M_\odot$. This is in a strong contrast with recent
claims that most SN Ia masses are peak around $1.4 M_\odot$
\citep{Scalzoetal2014}. \cite{Seitenzahletal2013} also claimed
that at least $50\%$ of all SN Ia come from near Chandrasekhar
mass ($M_{\rm Ch}$) WDs.

All in all, we conclude that the double-detonation scenario can
lead to explosions, but their characteristics are not typical of
those of SN Ia. Thus, SNe Ia must be originated  by other
channels, most likely the core-degenerate and the
double-degenerate scenarios \citep{TsebrenkoSoker2015b}.

{{{{ We thank an anonymous referee for many detailed comments
that substantially improved, both the presentation of
our results and their scientific content. }}}}
 This research was supported by the Asher Fund for Space
Research at the Technion, and the E. and J. Bishop Research Fund
at the Technion. This work was also partially supported by MCINN
grant AYA2011--23102, and by the European Union FEDER funds. OP is
supported by the Gutwirth Fellowship.

\end{document}